\documentclass[prb, twocolumn,  amsmath,amssymb]{revtex4-1}

\usepackage{amssymb}
\usepackage{amsmath}
\usepackage{ulem}
\usepackage{stackrel}
\usepackage{graphicx}

\usepackage{subfigure}
\usepackage{units}
\usepackage{color}

\usepackage{amssymb,color,ulem}
\usepackage{latexsym}
\usepackage{graphicx}
\usepackage{amsmath}
\usepackage{setspace}
\usepackage{fancyhdr}
\usepackage{graphics}
\usepackage{lastpage}

\usepackage{color}

\begin{document}

\title{Origin of spin dependent tunneling through chiral molecules}

\author{Karen Michaeli}
\affiliation{Department of Condensed Matter Physics, Weizmann Institute of Science, Rehovot 76100, Israel}
\author{Ron Naaman}
\affiliation{Department of Chemical Physics, Weizmann Institute of Science, Rehovot 76100, Israel}

\begin{abstract} 

The functionality of many biological systems depends on reliable electron transfer with minimal heating. Unlike man-made electric circuits, nature realizes electron transport via insulating chiral molecules. Here we include spin into the analysis of tunneling through these molecules, and demonstrate its importance for efficient transport. We show that the helical geometry induces robust spin filtering accompanied by, and intimately related to, strongly enhanced transmission. Thus, we resolve two key questions posed by transport measurements through organic molecules, demonstrating their common origin.
\end{abstract}

\maketitle

One of the main challenges in developing nano-scale electronic devices is minimizing heating. In silicon-based computer chips, dissipation is the primary obstacle to keeping up with Moore's law. Interestingly, electron transport via quantum tunneling, which is currently being explored as a route to reduce heating in semiconductor devices,\cite{Riel}  has long been implemented in nature.  In biological systems, electron transfer occurs via tunneling---direct or in several steps---through organic molecules,\cite{Gray,Blumberger} most of which exhibit a helical structure. The high efficiency of electron transfer in these systems, especially over distances of nanometers\cite{Winkler,Waldeck} and beyond,\cite{Yates} is unexpected for tunneling-based  transport and is one of the most pressing questions in the field. Recent experiments have revealed that transport through such helix-shaped molecules strongly depend on the electron's spin.\cite{Naaman2,Xie,Hamelbeck} Theoretical attempts to explain this effect\cite{Eremko,Mujica1,Mujica2,Mujica3,Gutierrez,Galperin} rely on large spin-orbit coupling, which is uncommon in organic materials. Here we show that the helical geometry induces correlations between the spin of the transferred electrons and their flow direction. In the tunneling regime, these connections can explain the large spin polarization measured in experiments  over an energy range of hundreds of meV, as well as the enhanced transmission through chiral molecules. The  directionality generated by the locking of the electron spin and momentum may hold the key to understanding the extremely low dissipation of electric transfer through organic molecules despite strong molecular vibrations.\cite{Budich} Suppressing heating effects is essential in biological systems as temperature changes of a few degrees cause denaturation---an irreversible structural deformation of proteins.

Organic molecules through which electrons flow may vary substantially in their composition, resistance and length. This article focuses on short helix shaped molecules, such as double-stranded DNA and oligopeptide, where the dependence of the conductivity on the length of the molecule is consistent with direct tunneling. However, the magnitude of the observed conductivity is much higher than anticipated based on first-principle calculations of the electronic states.\cite{Sek} Even more surprisingly, various experimental setups\cite{Xie,Hamelbeck}  find that over a wide energy range of hundreds of $meV$, the helicity of an electron, i.e., the projection of its spin onto its momentum direction, strongly affects the tunneling probability. This phenomenon, known as chiral-induced spin selectivity\cite{Naaman2} (CISS), is indicative of strong spin-orbit interaction, much larger than its expected value in organic materials.

Two important questions arising from the experimental observations are thus: why is the transmission through helical molecules in the tunneling regime larger than expected? And what causes the robust CISS? In this paper, we demonstrate that these two properties are strongly interlinked and propose a resolution to both questions. We show that while the spin-orbit coupling alone is too weak to account for the observations, in combination with the strong dipole electric field characteristic of these molecules, it can induce strong CISS. Moreover, we demonstrate that spin selectivity goes hand in hand with a dramatic enhancement of transmission through the molecule. To illustrate these properties, we construct a continuum effective model for electrons in the helical molecules that contains the minimal set of necessary ingredients. The first is a potential $V_H(\vec r)$ that confines electrons to propagate within a spiral tube\cite{Wang} centered around
\begin{align}\label{eq:HelixVector}
\vec{P}(s)=\hat{x}R\cos\left(\frac{2\pi s}{\tilde{R}}\right)+\hat{y}R\sin\left(\frac{2\pi s}{\tilde{R}}\right)+\hat{z}\left(\frac{b s}{\tilde{R}}\right),
\end{align}
as illustrated in Fig.~1(a). Here $s$ is the coordinate along the helix, $b$ is its pitch, $R$ its radius and the helix parameter $\tilde{R}=\pm\sqrt{(2\pi{R})^2+b^2}$ is positive (negative) for a right (left) handed spiral. $V_H$ models the periodic component of the potential generated by the atoms that comprise the molecule. The second ingredient is a dipole potential ${V}_{D}(\mathbf{r})$ that grows linearly along the central axis of the molecule. This field is a consequence of the dipolar nature of the hydrogen bonds, and in many cases also by  the amino acids terminating the molecule or by potential difference between the donor and the acceptor. Together, they give rise to a substantial voltage difference of $0.1-1$ Volt across the molecule. The dipole potential favors localization of electrons, and is the basis of the prediction that the tunneling conductivity should be significantly lower than measured. In the helical molecules, $V_{H}$ and $V_{D}$  both arise from the atomic potential, and they correspond to its periodic and non periodic components, respectively.

 \begin{figure}[t]
\begin{flushright}\begin{minipage}{0.5\textwidth} \centering
       \includegraphics[width=0.75\textwidth]{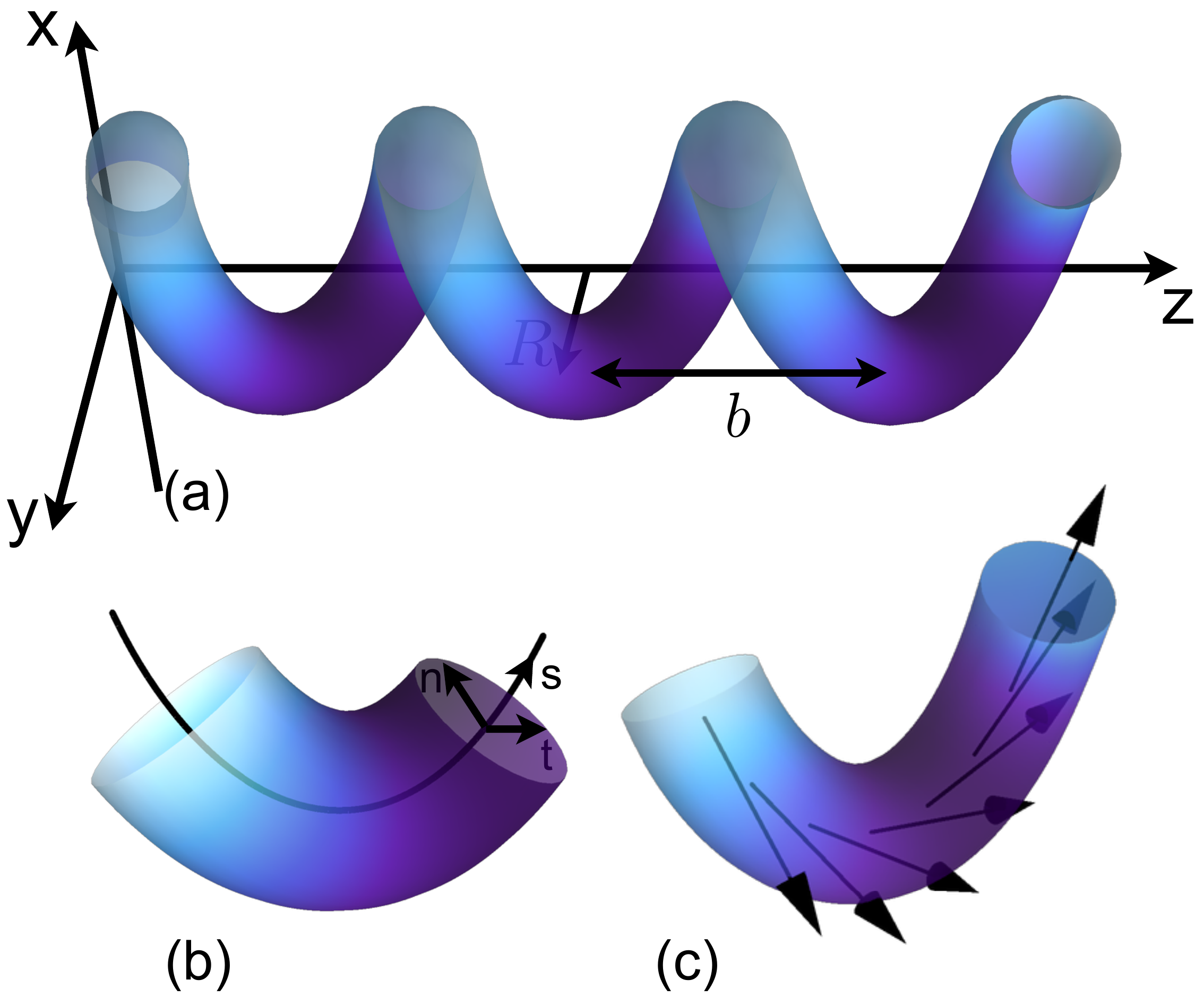}
                \caption[]{\small The Helix-shaped tube and the corresponding coordinate system. (a) Electrons are confined to within a helical tube of radius $R$ and pitch $b$.  (b) The Helical coordinate system is defined by the position along the spiral axis $s$ as well as $n$ and $t$ that span the perpendicular plane. (c) The spin orbit coupling in Eq.~\ref{eq:SOC} acts as an effective Zeeman field rotating as a function of position along the helix.}
\end{minipage}\end{flushright}
\end{figure}

The behavior of electrons in our model is governed by the single-particle Hamiltonian
\begin{align}\label{eq:Hamiltonian}
\mathcal{H}\hspace{-0.5mm}=\hspace{-0.5mm}-\frac{\hbar^2\mathbf{\nabla}^2}{2m}+V_{H}(\mathbf{r})+{V}_{D}(\mathbf{r})+\hspace{-0.5mm}\frac{i\hbar^2}{4m^2c^2}\mathbf{\sigma}\hspace{-0.5mm}\cdot\hspace{-0.5mm}\left(\mathbf{\nabla}V_{H}(\mathbf{r})\hspace{-0.5mm}\times\hspace{-0.5mm}\mathbf{\nabla}\right)\hspace{-0.5mm}.
\end{align}% 
 The last term in the above Hamiltonian is the spin-orbit coupling which arises as a leading relativistic correction to the Schr{\"o}dinger equation.  The precise form of $V_H$ is not crucial for our analysis; for specificity we here take it to be an isotropic harmonic potential in the plane perpendicular to the helix axis. Thus, $V_{H}(\mathbf{r})\approx\frac{\hbar^2\rho^2}{2ma_0^4}$ where $(\rho,\theta)$ are the spherical coordinates in this plane and $a_0 \ll \tilde R$ is the radius of the tube. To lowest order in $a_0/\tilde R$, each eigenstate can be written as a product of an $s$-dependent function and the wave-function of a two dimensional harmonic oscillator in the $(\rho,\theta)$ plane. The latter are labeled by the level index $N\in \mathbb{N}$ and $\ell=-N,-N+2...N-2,N$. The quantum number $\ell$ denotes the eigenvalues of $-id/d\theta$, i.e., the angular momentum operator pointing perpendicular to the plane.  For the helix-shaped cylinder this vector $\vec L_\text{helix}$ is tangent to the helix vector $\vec{P}$ defined in Eq.~\ref{eq:HelixVector}, and hence changes as function of position $s$.

\begin{figure}[ht]
 \begin{flushright}\begin{minipage}{0.5\textwidth} \centering
       \includegraphics[width=0.9\textwidth]{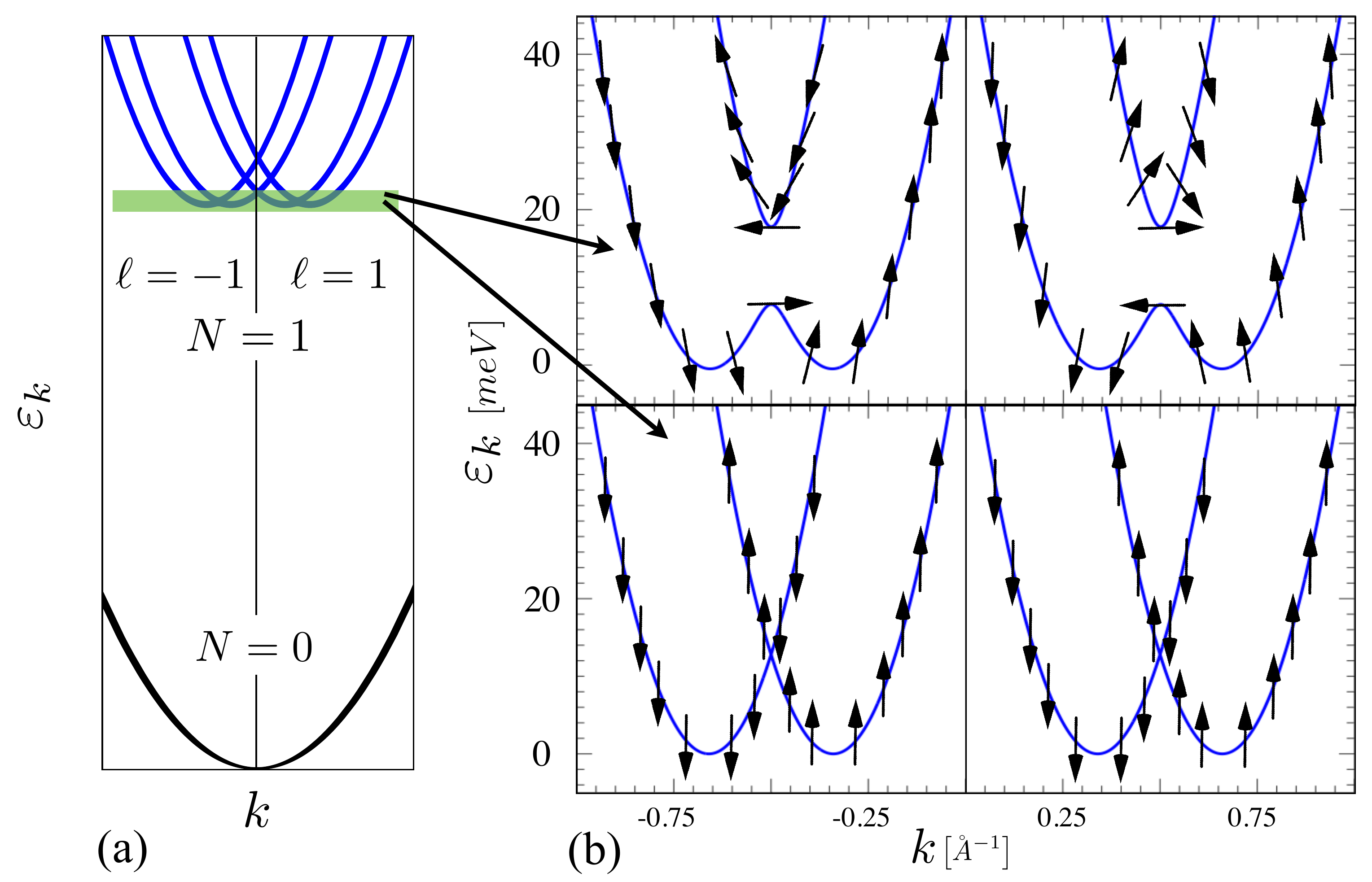}
\caption[]{\small  Electronic band structure for $V_{D}=0$. (a) The $N=1$ bands, schematically shown as function of linear momentum $k$ along the helix axis, are split by a large energy difference from the $N=0$ band. (b) The exact spectrum for the $N=1$ band with (upper) and without (lower) spin-orbit coupling features a partial gap opening for $\kappa\neq0$. For the derivation we assumed $b=R=0.3nm$ as well as a delocalized band width of $\hbar^2/2m\sim1eV\AA^2$. The arrows indicate the spin direction of the electronic state.}
\end{minipage}\end{flushright}
\end{figure}

 \begin{figure*}[tp]
  \begin{flushright}\begin{minipage}{1\textwidth} \centering
       \includegraphics[width=0.95\textwidth]{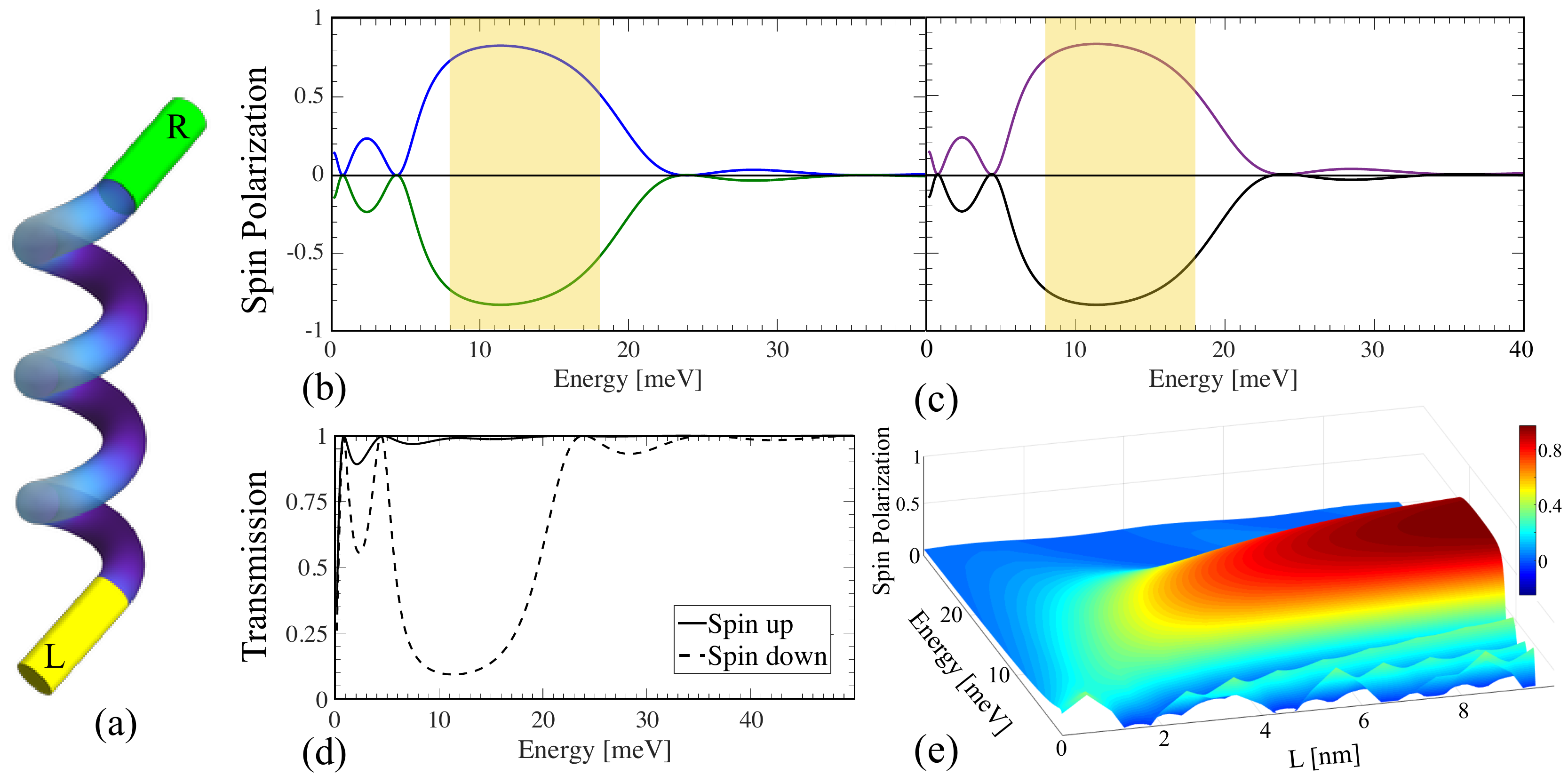}
                \caption[]{\small Transmission through helix-shaped molecule without a dipole field. (a) The setup considered in the derivation of the transmission probability assumes an helical molecule attached to two straight cylindrical leads with the same parameters as the spiral system. Different realizations are discussed in Appendix B. The spin polarization -- the difference between a the transmission probabilities of a polarized current with spin up and spin down (normalized by the total transmission)-- as a function of energy of incoming electrons with $N=1$ is shown in panel (b) and (c). The upper curves correspond to transmission of electrons with  $\ell=1$ (blue), $\ell=-1$ (purple)  between the left and right right leads through a right handed molecule. The lower curves illustrate that the opposite polarization is obtained when a left handed molecule is considered (green) or the current is sent from right to left (black). The shaded regions mark the energies of the partial gap. (d) The  transmission per spin clearly shows the reduced probability for one helicity compared to the other. (e) The achievable polarization increases with the length of the molecule.}
\end{minipage}\end{flushright}
 \end{figure*}

The coupling between the angular momentum and $s$ due to the helical geometry becomes evident when the Hamiltonian is expanded to order $\left(\nicefrac{\displaystyle a_0}{ \tilde{R}}\right)^2$:
\begin{align}\label{eq:Hamiltonian2}
\mathcal{H}_{N,\ell}(s)&=E_{N}+V_D(s)-\frac{\hbar^2}{2m}\frac{\partial^2}{\partial{s}^2}+i\frac{\hbar^2\gamma\ell}{m}\frac{\partial}{\partial{s}}\\\nonumber&+
\kappa\vec{\sigma}\cdot\vec{L}_{\text{helix}}+\Delta{E}_{\ell}.
\end{align}% 
Here we  introduced the parameters $\gamma=\frac{b}{\tilde{R}^2}$ and $\kappa=\frac{\hbar^4R}{4m^3a_0^4c^2\tilde{R}}$ that are smaller by a factor of $\left(\nicefrac{\displaystyle a_0}{ \tilde{R}}\right)^2$ than the leading contribution $E_{N}=\frac{\hbar^2}{ma_0^2}(N+1)$ (see Appendix A  for details). The first three terms describe electrons in a (straight) cylinder with a potential $V_D(s)=eE_Ds$ that increases linearly along the central axis. The remaining terms are unique to this problem, and reflect the helical geometry. The last term is a trivial shift of the energy levels $\Delta{E}=\frac{\hbar^2b^2\ell^2}{2m\tilde{R}^4}$. The physical interpretation of the other two terms is also clear: $i\gamma\ell\partial_s$ describes the centripetal potential felt by a particle moving along the helix. It corresponds to the classical effect that a particle propagating through a spiral pipe gets pushed further up along the sides of the tube at higher velocity. The final term has the familiar form of spin-orbit coupling due to a confining potential. For a helical system, as shown in Fig.~1(c), it resembles a Zeeman magnetic field with a component that rotates in the $x-y$ plane as a function of position along the helix:
\begin{align}\label{eq:SOC}
\kappa\vec{\sigma}\cdot\vec{L}_{\text{helix}}=\kappa\ell\left[\sigma_x\sin\frac{2\pi{s}}{\tilde{R}}-\sigma_y\cos\frac{2\pi{s}}{\tilde{R}}-\sigma_z\frac{b}{2\pi{R}}\right].
\end{align}%
Crucially, sign and magnitude of the field are determined by the angular momentum $\ell$, and hence this term preserves time-reversal symmetry. Since, the constant component of the spin-orbit term along the $z$-direction has a negligible effect on the spin-dependent transport, we will suppress it from here on. Below we show that the rotating component of the magnetic field gives rise to the CISS effect. The Hamiltonian can be further simplified by the  unitary transformation  $\Psi_{\sigma}\rightarrow{e^{2\pi{i}\sigma{s}/\tilde{R}}}\bar{\Psi}_{\sigma}$, where $\sigma=\pm$ denotes spin pointing along the $\pm{z}$-direction. Under this transformation the Hamiltonian becomes:
\begin{align}\label{eq:Hamiltonian3}
\mathcal{H}_{N,\ell}(s)&=E_{N}+V_D(s)-\frac{\hbar^2}{2m}\left(\frac{\partial}{\partial{s}}-i\frac{\pi}{\tilde{R}}\sigma_z\right)^2\\\nonumber
&+i\frac{\hbar^2\gamma\ell}{m}\left(\frac{\partial}{\partial{s}}-i\frac{\pi}{\tilde{R}}\sigma_z\right)+
\kappa\ell\sigma_y+\Delta{E}_{\ell}.
\end{align}% 
The rotating field has been transformed into a Rashba-like spin-orbit coupling and a constant Zeeman field.\cite{Loss} One-dimensional electronic systems described by similar Hamiltonians have attracted a lot of attention in recent years in by the condensed matter community as a platform for engineering topological superconductors.\cite{Oreg}

\begin{figure*}[t]
 \begin{flushright}\begin{minipage}{1\textwidth} \centering
       \includegraphics[width=0.95\textwidth]{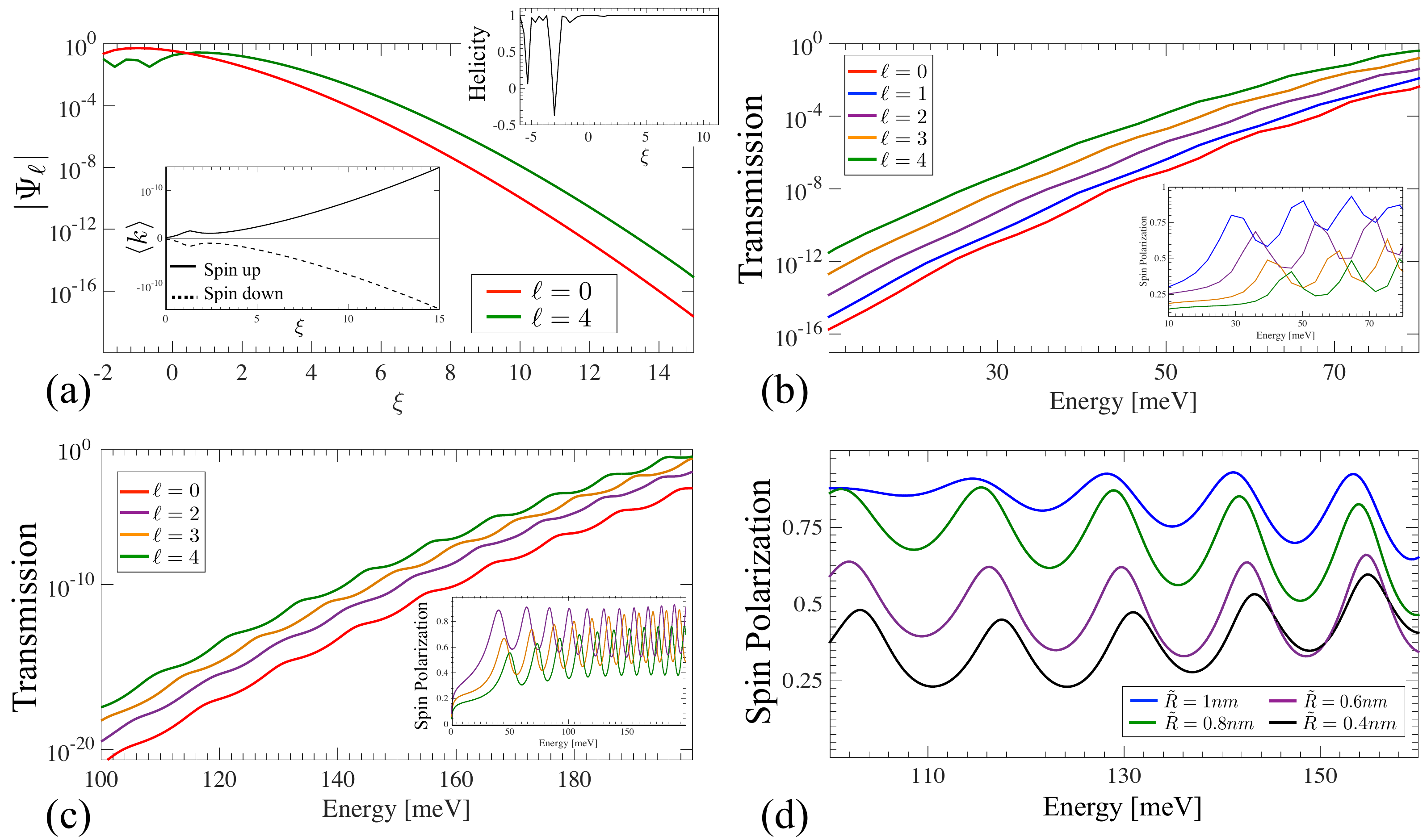}
                \caption[]{\small Transmission through an helix-shaped molecule in the presence of a dipole field. (a) In the presence of spin-orbit coupling, the amplitude of the (exact) electronic wave-function in the tail $\xi\gg1$ grows as a function of the angular momentum.  Moreover, the spin is aligned along the momentum direction (see inset), and as a consequence the state has a well defined helicity. (b) The increased amplitude deep inside the molecule give rises to an enhanced transmission probability that grows with $\ell$.  The scattering matrix\cite{Entin} is derived for the exact wave-function (see Appendix C), with $L=2.15nm$, $R=b=7.5nm$, $\kappa=5meV$, $E_D=5\cdot10^{7}eV/m$ and  assumed a narrower band, i.e., an effective mass in the $s$-direction that is about 50 times larger than the free electron mass. This panel shows that the enhanced transmission for $\ell\neq0$ is accompanied by a spin polarization (inset) of order unity over an energy range exceeding $\kappa\ell$. Similar results were obtained using a tight binding calculation (c) for a molecule with the same parameter but somewhat different length $L=5nm$. (d) Deforming the molecule to increase $\tilde{R}$ (through a larger pitch or radius) helps  spin polarization. This observation holds as long as $\eta_0\sqrt{\nicefrac{\eta_0}{L}}\gg\tilde{R}\ll{L}$.}
\end{minipage}\end{flushright}
\end{figure*}

Without the dipole field, $V_D=0$, the electronic spectrum consists of  bands, labeled by $N$, $\ell$ that disperse with momentum $k$ along $s$. As demonstrated in Fig.~2(a), the Rashba-like term splits the energy spectrum of spin-up and down electrons, while the Zeeman field opens a partial gap at energies $\Delta_{soc}\pm|\kappa\ell|=\frac{\pi^2\hbar^2}{2m\tilde{R}^2}\pm |\kappa\ell |$. Typical parameters for organic molecules, $R,b\approx{0.3nm}$ and $\kappa=5meV$, correspond to $\Delta_{soc}/|\kappa\ell|\approx10$ for $\ell=1$, where a delocalized band with the free electron mass $m$ is assumed. For $|\kappa\ell|<\Delta_{soc}$, the states within the partial gap are quasi-helical; the spin is almost perfectly locked to the momentum direction as shown in Fig.~2(b), resulting in CISS. We studied the spin-dependent scattering matrix\cite{Entin} from which we extract the probability $T_{N,\ell,\sigma}(\epsilon)$ for an electron injected with spin $\sigma$ and energy $\epsilon$ (as well as quantum number $N$ and $\ell$) to be transmitted through a molecule of length $L$.  The result of the calculation is summarized in Fig.~3. We find that spin polarization, $\mathcal{P}=\frac{T_{\uparrow}-T_{\downarrow}}{T_{\uparrow}+T_{\downarrow}}$, becomes of order unity at energies within the partial gap, with the sign of $\mathcal{P}$ independent of the sign of $\ell$. Due to the helical nature of the electronic state in this energy window, the polarization is reversed for transmission in the opposite direction. Moreover, the helicity of electronic states inside the partial gap, and correspondingly the polarization direction, is determined by the handedness of the helix (the sign of the Rashba-like term).

We have seen that the helical geometry naturally gives rise to spin dependent transport. Yet, for molecules with delocalized electronic states it can account for CISS only in a narrow energy windows of $\sim 1meV$.  Previous theoretical studies invoked  properties of the leads,\cite{Nitzan} next nearest neighbor hopping\cite{Guo,Cuniberti} or high density of molecules\cite{Mujica2,Kuzmin,Vager} to explain the strong signatures.  Here we wish to emphasize that molecules that exhibit significant CISS also feature a large dipole potential $V_D(s)=eE_Ds$ which was absent in the previous analyses; we now turn to discuss the effects of this term. In the absence of the unique spin-orbit term (Eq.~\ref{eq:SOC}), or for bands with $\ell=0$, the decay of an electron wave-function of energy $\varepsilon$ into the potential barrier is described by the Airy function $Ai(\xi)\xrightarrow[\xi\rightarrow\infty]{}\frac{\exp\left(-\nicefrac{2\xi^{3/2}}{3}\right)}{2\sqrt{\pi}\xi^{1/4}} $. Here we introduced the dimensionless coordinate $\xi = s/\eta_0-\varepsilon/\mathcal{E}$, where $\eta_0=\left(\frac{\hbar^2}{2mE_D}\right)^{1/3}$  and $\mathcal{E}=\left(\frac{\hbar^2E_D^2}{2m}\right)^{1/3}$ are the characteristic length and energy scales, respectively. Eigenstates  with non-zero $\ell$ exhibit a non-trivial spin dependence:  the exact wave-functions of the Hamiltonian in Eq.~\ref{eq:Hamiltonian3} with arbitrary $\ell$ are of the form
\begin{align}\label{eq:EigenStates1}
\Psi_{\ell}^{(1)}(\xi)=\left(\begin{array}{c}
\mathcal{U}_{\ell}(\xi) \\
\mathcal{V}_{\ell}(\xi)\end{array}\right)\hspace{10mm}
\Psi_{\ell}^{(2)}(\xi)=\left(\begin{array}{c}
\mathcal{V}_{\ell}^{*}(\xi) \\
\mathcal{U}_{\ell}^{*}(\xi)\end{array}\right).
\end{align}
The derivation and exact expressions of the functions $\mathcal{U}_{\ell}(\xi)$ and $\mathcal{V}_{\ell}(\xi)$ are given in Appendix C.  Here we focus on their amplitude deep inside the molecule: 
\begin{align}\label{eq:Asymptotics2}
&|\mathcal{U}_{\ell}(\xi)|\hspace{-0.5mm}\xrightarrow[\xi\gg\eta_0^2/\tilde{R}^2,\tilde{R}/\eta_0
]{}\hspace{-0.5mm}\frac{1}{2}\hspace{-0.5mm}\left[\exp\hspace{-0.5mm}\left(\frac{\kappa^2\ell^2\tilde{R}}{4\eta_0\mathcal{E}^2}\right)\hspace{-1mm}+\hspace{-0.5mm}1\hspace{-0.5mm}\right]\hspace{-0.5mm}Ai(\xi); \\\nonumber
&|\mathcal{V}_{\ell}(\xi)|\hspace{-0.5mm}\xrightarrow[\xi\gg\eta_0^2/\tilde{R}^2,\tilde{R}/\eta_0]{}\hspace{-0.5mm}\frac{\kappa\ell}{2\eta_0\mathcal{E}\sqrt{\xi}}\hspace{-0.5mm}\left[\exp\hspace{-0.5mm}\left(\frac{\kappa^2\ell^2\tilde{R}}{4\eta_0\mathcal{E}^2}\right)+\hspace{-0.5mm}1\hspace{-0.5mm}\right]\hspace{-0.5mm}Ai(\xi).
\end{align}

From the asymptotic expansion of the wave-functions we obtain two important properties of the helical molecules that are the main results of this work. Firstly, for $\frac{\kappa^2\ell^2\tilde{R}}{4\eta_0\mathcal{E}^2}\gg1$ and $\eta_0\sqrt{\nicefrac{\eta_0}{L}}<\tilde{R}<L$ the amplitude of the wave-functions at the end of the molecule is significantly larger than for $\ell=0$ or $\kappa=0$. Figure 4 shows this property as well as the enhanced transmission probability for $\ell\neq0$ based on the exact wave-functions in the continuum model, and from a numerical tight-binding\cite{Lee} calculation (see Appendix C for details).  Secondly, the wave-functions $\Psi_{\varepsilon,\ell}^{(1)}(s)$ and $\Psi_{\varepsilon,\ell}^{(2)}(s)$ have (the same) well defined helicity in their tail that gives rise to CISS of the same sign as in the $V_{D}=0$ case (see Fig.~4). The energy range where strong spin-dependent transport can be observed, however, is determined by $eE_DL\sim0.1eV$ instead of by the partial gap $\kappa\ell \sim 1meV$. Thus, although the dipole field reduces the total transmission through the molecule, over a wide range of energies it preferentially transmits electrons with one helicity. One way to understand this behavior is via its connection to Klein tunneling\cite{Klein} through a linearly increasing potential.\cite{Sauter} Such a barrier is transparent for massless relativistic Dirac particles, while for massive particles the transmission through the barrier gets exponentially small. From the energy spectrum shown in Fig.~2(b) we see that states in the vicinity of the gap are effectively described by a massive Dirac equation. Thus, for $\frac{m\kappa^2\ell^2\tilde{R}}{2\hbar^2\pi{E}_D^2}\gg1$ these states get strongly localized by the potential. In contrast, the scattering rate between these modes and high momentum states get suppressed.\cite{Brouwer} As a result, the penetration of the high momentum, quasi-helical states into the barrier is enhanced.

The strong spin dependence of electron transmission through helical organic molecules is by now experimentally well established. We showed that enhanced transmission and CISS \textit{both} originate in the interplay of spin-orbit coupling and a dipolar potential.  The helical nature of the wave-function obtained here suggests that backscattering by disorder or phonon scattering\cite{Budich} is suppressed, and hence, the conductivity through the molecule is robust. Our model provides a simple platform for future studies of these effects, as well as clear predictions (see Fig.~4(d)) how conductivity through helical molecules depends on various parameters. \\

\begin{acknowledgements}
We thank Y. Imry  for stimulating conversations and important insights. We also gratefully acknowledge support from the Gurwin Family Fund for Scientific research (KM); and from an ERC-Adv grant (RN).
\end{acknowledgements}

\appendix

\onecolumngrid

\section{Derivation of the Hamiltonian}

The starting point for the derivation of the effective Schr{\"o}dinger equation for electrons in an helix-shaped tube is the Hamiltonian given in Eq.~2. This Hamiltonian accounts for the kinetic energy in three dimensional space $\mathcal{H}_k=-\frac{\hbar^2\mathbf{\nabla}^2}{2m}$, the potential $V_H$ that confines the electrons to the spiral tube, the dipole field $V_D$ that is approximately growing linearly with $z$, and the spin-orbit coupling $\mathcal{H}_{soc}$.  The first step in the derivation is to transform into the helical coordinate system defined by the vectors:~\cite{Wang}
\begin{align}\label{S1}\nonumber
\tilde{R}\hat{Q}&=-{2\pi{R}}\sin\left({2\pi{s}}/{\tilde{R}}\right)\hat{x}+{2\pi{R}}\cos\left({2\pi{s}}/{\tilde{R}}\right)\hat{y}+{b}\hat{z};\\
\hat{N}&=-\cos\left({2\pi{s}}/{\tilde{R}}\right)\hat{x}-\sin\left({2\pi{s}}/{\tilde{R}}\right)\hat{y};\\\nonumber
\tilde{R}\hat{T}&=-{b}\sin\left({2\pi{s}}/{\tilde{R}}\right)\hat{x}+{b}\cos\left({2\pi{s}}/{\tilde{R}}\right)\hat{y}-{2\pi{R}}\hat{z}.
\end{align}
The vector $\hat{Q}$ is tangential to the curve $\vec{P}(s)$ defining the helix axis, while  $\hat{N}$ and $\hat{T}$ span the perpendicular plane, see Fig.~1 (b). Correspondingly, we define the coordinates $s$, $\rho$ and $\theta$, where the first denotes the position along the helix, while the other two are the spherical coordinates in the $\hat{N}-\hat{T}$ plane.  In the helix basis defined above, the cartesian coordinates become:
\begin{align}\label{S2}\nonumber
x&=\left(R-\rho\cos\theta\right)\cos\left({2\pi{s}}/{\tilde{R}}\right)+\frac{b\rho}{\tilde{R}}\sin\theta\sin\left({2\pi{s}}/{\tilde{R}}\right);\\\nonumber
y&=\left(R-\rho\cos\theta\right)\sin\left({2\pi{s}}/{\tilde{R}}\right)-\frac{b\rho}{\tilde{R}}\sin\theta\cos\left({2\pi{s}}/{\tilde{R}}\right);\\
z&=\frac{bs}{\tilde{R}}+\frac{2\pi{R}\rho}{\tilde{R}}\sin\theta.
\end{align}%
In this coordinate system, the potential $V_{H}$ is a function of $\rho$ alone. We approximate it by the harmonic potential $V_{H}(\rho)=\frac{\hbar^2\rho^2}{2ma_0^4}$. For $a_0\sim0.5\AA$, this potential creates energy bands separated by a gap of order  $0.5eV-1eV$.

Applying the coordinates transformation to the helix basis, the kinetic term acquires the following form:
\begin{align}\label{S3}
\mathcal{H}_{k}&=-\frac{\hbar^2}{2m}\left\{\frac{\partial^2}{\partial\rho^2}+\left[\frac{1}{\rho}-\frac{2\pi{R}\cos\theta}{\tilde{R}^2-2\pi{R}\rho\cos\theta}\right]\frac{\partial}{\partial\rho}+\left[\frac{1}{\rho^2}+\frac{b^2}{(\tilde{R}^2-2\pi{R}\rho\cos\theta)^2}\right]\frac{\partial^2}{\partial\theta^2}\right.\\\nonumber
&\left.+\frac{2\pi{R}\sin\theta}{\rho(\tilde{R}^2-2\pi{R}\rho\cos\theta)}\left[1-\frac{b^2\rho^2}{(\tilde{R}^2-2\pi{R}\rho\cos\theta)^2}\right]\frac{\partial}{\partial\theta}+\frac{\tilde{R}^4}{(\tilde{R}^2-2\pi{R}\rho\cos\theta)^2}\frac{\partial^2}{\partial{s}^2}\right.\\\nonumber
&\left.+\frac{2b\tilde{R}^2}{(\tilde{R}^2-2\pi{R}\rho\cos\theta)^2}\frac{\partial^2}{\partial{s}\partial\theta}+\frac{bR\tilde{R}^2\rho\sin\theta}{(\tilde{R}^2-2\pi{R}\rho\cos\theta)^3}\frac{\partial}{\partial{s}}\right\}.
\end{align}%
The rather complicated expression for the kinetic energy is a consequence of the fact that motion along the helix axis corresponds to a translation in the $z$ direction combined with rotation around it. Thus, in the helical basis the kinetic energy intertwines the coordinate $s$ with the degrees of freedom  in the perpendicular direction $\rho$ and $\theta$. Similarly, we can write the dipole potential as $V_D=E_D\left[s+\frac{2\pi{R}\rho}{b}\sin\theta\right]$.
Here we define the electric field $E_D$ as the slope of the dipole potential along the helix axis. Since the dipole field is directed along the $z$-axis, $E_D$ is smaller by a factor of $b/\tilde{R}$. Finally the spin-orbit coupling becomes:
\begin{align}\label{S5}\nonumber
\mathcal{H}_{soc}&=i\frac{\hbar^2}{4m^2c^2}\frac{\tilde{R}}{\rho(\tilde{R}^2-R\rho\cos\theta)}\frac{\partial{V_{H}}}{\partial\rho}\left\{\frac{b}{2\pi}\sigma_z\hspace{-1mm}-\hspace{-1mm}\left[(R-\rho\cos\theta)\sin\left(\frac{s}{\tilde{R}}\right)-\frac{b\rho}{2\pi\tilde{R}}\sin\theta\cos\left(\frac{s}{2\pi\tilde{R}}\right)\right]\sigma_x\right.\\
&\left.+\left[(R-\rho\cos\theta)\cos\left(\frac{s}{\tilde{R}}\right)+\frac{b\rho}{2\pi\tilde{R}}\sin\theta\sin\left(\frac{s}{2\pi\tilde{R}}\right)\right]\sigma_y\right\}\frac{\partial}{\partial\theta}\\\nonumber&+\frac{\tilde{R}}{(\tilde{R}^2-R\rho\cos\theta)}\frac{\partial{V_{H}}}{\partial\rho}\left\{R\cos\theta\sigma_z-\left[\frac{b}{2\pi}
\cos\theta\sin\left(\frac{s}{\tilde{R}}\right)+\tilde{R}\sin\theta\cos\left(\frac{s}{2\pi\tilde{R}}\right)\right]\sigma_x\right.\\\nonumber
&\left.+\left[\frac{b}{2\pi}
\cos\theta\cos\left(\frac{s}{\tilde{R}}\right)-\tilde{R}\sin\theta\sin\left(\frac{s}{2\pi\tilde{R}}\right)\right]\sigma_y\right\}\frac{\partial}{\partial{s}}.
\end{align}%
The spin-orbit term contains two main contributions,  one that couples the spin to the angular momentum $\ell$, and second that connect the spin to the linear momentum along the helix.~\cite{Jaffe} In the derivation of the spin-orbit coupling term we neglect the effect of the dipole potential, since it is substantially smaller than the contribution due to the helix potential $V_H$.

The transformation into the helical coordinate system simplifies the search for electronic states when the radius of the tube $a$ is much smaller than the helix parameter $\tilde{R}$. In other words, as long as the tube is substantially narrower than the radius or pitch of the spiral. This limit, which is realized in organic helical polymers, allows us to expand the Hamiltonian in orders of $a_0/\tilde{R}$.  At leading order, the Hamiltonian separates into a $s$-dependent part and a $(\rho,\theta)$-dependent part:
\begin{align}\label{S6}
\mathcal{H}_{0}&=-\frac{1}{2m}\left[\frac{\partial^2}{\partial\rho^2}+\frac{1}{\rho}\frac{\partial}{\partial\rho}+\frac{1}{\rho^2}\frac{\partial^2}{\partial\theta^2}\right]+\frac{\hbar^2\rho^2}{2ma_0^4}-\frac{\hbar^2}{2m}\frac{\partial^2}{\partial{s}^2}+V_D(s).
\end{align}%
The eigenstates of this Hamiltonian can be written as a product of an $s$-dependent function $\Psi(s)$ and the wave-functions of a two dimensional  harmonic oscillator $\Phi_{N,\ell}\left(\rho,\theta\right)=\sqrt{\frac{n_r!}{\pi{a}_0^2(n_r+|\ell|)!}}e^{i\ell\theta-\rho^2/2a_0^2}$ $\left(\frac{r}{a_0}\right)^{|\ell|}L_{n_r}^{|\ell|}\left(\frac{r^2}{a_0^2}\right)$, where $L_{m}^{\nu}(x)$ is the generalized Laguerre polynomial, and $N=2n_r+|\ell|$. The corresponding energy levels of the harmonic oscillator $E_N=\frac{\hbar^2}{2ma_0^2}(N+1)$ are independent of $\ell=-N,-N+2...N-2,N$. To obtain next order corrections to the Hamiltonian, we project it onto the $N$th energy level: 
\begin{align}\label{S7}
\langle\Phi_{N,\ell}|\mathcal{H}|\Phi_{N,\ell'}\rangle=\int{\rho\left(1-\frac{R\rho}{\tilde{R}}\cos\theta\right)d\rho{d}\theta}\Phi_{N,\ell}\left(\rho,\theta\right)\mathcal{H}\Phi_{N,\ell'}\left(\rho,\theta\right).
\end{align}
The factor $\rho(1-R\rho\cos\theta/\tilde{R})$ is the measure of the integral in the helical coordinate system. In this approximation we ignore transitions between states with different $N$ as they give rise to corrections of order $(a_0/\tilde{R})^3\ll1$.  Keeping only corrections of order $(a_0/\tilde{R})^2$, the Hamiltonian reduces to Eq.~3. Since the first order corrections to the Hamiltonian already couple the angular momentum and the coordinate along the helix axis, the $s$ dependent part of the wave-function becomes a function of $\ell$ well, $\Psi(s)\rightarrow\Psi_{\ell}(s)$.

\section{Electronic states and spin selectivity in the metallic limit}

In the absence of a dipole field all electronic states are extended. The Hamiltonian in Eq.~5 with $V_{D}=0$ is diagonal in momentum space, and the Schr{\"o}dinger equation for $\Psi_{\ell}(k)=\int{ds}e^{iks}\Psi_{\ell}(s)$ becomes:
 \begin{align}\label{S8}
\left[\frac{\hbar^2}{2m}\left(k-\frac{\pi}{\tilde{R}}\sigma_z\right)^2\hspace{-1.5mm}+\hspace{-0.5mm}
\kappa\ell{\sigma}_y\right]\Psi_{\ell}(k)=E_{N,\ell}(k)\Psi_{\ell}(k).
\end{align}%        
Here $k=-i\partial/\partial{s}-\gamma\ell$ is proportional to the the momentum along the helix axis. The corresponding wave-functions are:
\begin{align}\label{S9} 
\Psi_{\ell}^{+}(k)=\frac{1}{\sqrt{2}}\left({\setstretch{2.5}\begin{array}{c}
i\sqrt{1+\frac{k}{\sqrt{k^2+4\tilde{R}^2\kappa^2\ell^2/\pi^2}}} \\
\sqrt{1-\frac{k}{\sqrt{k^2+4\tilde{R}^2\kappa^2\ell^2/\pi^2}}}\end{array}}\right); \hspace{5mm}
\Psi_{\ell}^{-}(k)=\frac{1}{\sqrt{2}}\left({\setstretch{2.5}\begin{array}{c}
-i\sqrt{1-\frac{k}{\sqrt{k^2+4\tilde{R}^2\kappa^2\ell^2/\pi^2}}} \\
\sqrt{1+\frac{k}{\sqrt{k^2+4\tilde{R}^2\kappa^2\ell^2/\pi^2}}}\end{array}}\right),
\end{align}
with energies  $E_{N,\ell}^{\pm}(k)=E_{N}+\Delta{E}_{\ell}-\frac{\hbar^2\gamma^2\ell^2}{2m}+\frac{\hbar^2\pi^2}{2m\tilde{R}^2}+\frac{\hbar^2k^2}{2m}\pm\sqrt{\left(\frac{\pi\hbar^2k}{2m\tilde{R}}\right)^2+\kappa^2\ell^2}$.

\begin{figure}[tp]
\begin{flushright}\begin{minipage}{1\textwidth}  \centering
     \includegraphics[width=1\textwidth]{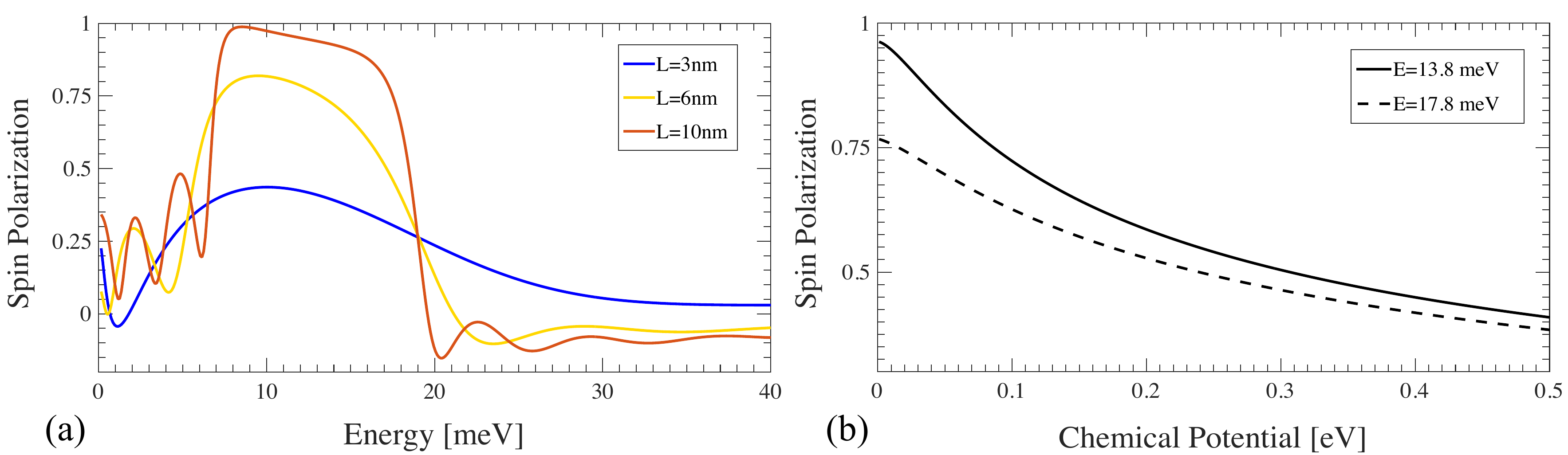} \hspace{0.05in}
              \caption[0.4\textwidth]{\small Transmission through a metallic helix-shaped molecule. (a) Strong polarization along the $z$-direction is obtained for an incident current that is spin polarized along the $x$-direction. (b) For energies within the partial gap ($2\kappa\ell=10meV$), changing the chemical potential in the lead by $500meV$ reduces the polarization only by a factor of  2-3. }
\end{minipage}\end{flushright}
\end{figure}

To find the spin dependent transmission coefficients,~\cite{Entin} we consider the setup illustrated in Fig.~3; the molecule (spiral tube) is connected to straight cylindrical leads of the same radius $a_0$. Thus, the only difference between the molecule and the leads is in the curvature effects that are absent in the latter. Correspondingly, the electronic states in the leads are characterized by the same quantum numbers $N$, $\ell$ and $k$, however, the energy spectrum $E_{N,\ell}^{\uparrow,\downarrow}(k)=E_{N}+\frac{\hbar^2k^2}{2m}$ is spin degenerate. The scattering matrix is found by matching the boundary condition at the points where the molecule is connected to the leads. Above we show the transmission probability for incident currents with spin polarized along the $\pm{z}$-direction . We demonstrate there that within the partial gap one helicity has higher transmission probability than the other. For completeness, we show in Fig.~5(a) that for incident current polarized in a $x$-direction, the outgoing states are nevertheless spin polarized along the $z$-direction. Moreover, so far we assumed that the bottom of the electronic bands in the molecule and the leads coincide. In Fig.~5(b), we show that changing the chemical potential of the leads has a relatively small effect.

\section{Electronic states and spin selectivity in the presence of a dipole field}

The large dipole potential characterizing the molecule gives us a unique opportunity to consider the effect of an electric field (or linear modulations of the chemical potential) on electronic states in the presence of spin-orbit coupling. In condensed matter systems such as semiconducting wires, in which the Hamiltonian Eq.~5 can be realized, a non-zero dipole potential $V_D\neq0$ will be screened.  The large energy gap  of proteins used for electronic transfer in biological systems, in contrast, prevents significant screening effects. To find the electronic state, we start with the Hamiltonian in Eq.~5 with $V_D=E_Ds$. In the following derivation, we assume a positive dipole field $E_D>0$. Solutions for $E_D<0$ are obtained by replacing $s\rightarrow-s$ and $\tilde{R}\rightarrow-\tilde{R}$.

Defining the length and energy scales  $\eta_0=\left(\frac{\hbar^2}{2mE_D}\right)^{1/3}$, $\mathcal{E}=\left(\frac{\hbar^2E_D^2}{2m}\right)^{1/3}$, the corresponding Schr{\"o}dinger equation becomes:
\begin{align}\label{S10}
\left[-\left(\frac{\partial}{\partial{\xi}}-i\eta_0\gamma\ell-i\frac{\pi\eta_0}{\tilde{R}}\sigma_z\right)^2+\xi+
\frac{\kappa\ell}{\mathcal{E}}\sigma_y\right]\Psi_{\ell}(\xi)=0.
\end{align}% 
Here, we use the dimensionless coordinate $\xi = s/\eta_0-\varepsilon/\mathcal{E}$, and the energy  is defined with respect to the bottom of the band  $\varepsilon=E-E_N-\Delta{E}_{\ell}+\frac{\hbar^2\gamma^2\ell^2}{2m}$. The eigenstates of the above Hamiltonian can be written as a convolution of the Airy functions $Ai(\xi)$ or $Bi(\xi)$ and a spinor:
\begin{align}\label{S11}
\Psi_{\ell}(\xi)=\hspace{-0.5mm}\left(\begin{array}{c}
\mathcal{U}_{\ell}(\xi) \\
\mathcal{V}_{\ell}(\xi) \end{array}\right)\hspace{-0.5mm}=
\hspace{-0.5mm}e^{i\gamma\eta_0\ell\xi}\hspace{-0.5mm}\int\hspace{-0.5mm}{d\chi}\hspace{-0.5mm}\left(\begin{array}{c}
f_{\ell}(\xi-\chi) \\
g_{\ell}(\xi-\chi)\end{array}\right)Y(\chi).
\end{align} 
For $Y(\xi)=Ai(\xi)$ ( $Y(\xi)=Bi(\xi)$) the wave-function decays (diverges) at large $\xi$. Interestingly, the Schr{\"o}dinger equation for the spinor is identical to the Dirac equation for a massive particle in the presence of linear potential:~\cite{Sauter}
\begin{align}\label{S12}
\left(\hspace{-1mm}\begin{array}{cc}
i\frac{2\pi\eta_0}{\tilde{R}}\partial_{\xi}+\xi+\frac{\pi^2\eta_0^2}{\tilde{R}^2\mathcal{E}} & \hspace{-3mm} -i\frac{\kappa\ell}{\mathcal{E}}\\
i\frac{\kappa\ell}{\mathcal{E}} &  \hspace{-3mm}
-i\frac{2\pi\eta_0}{\tilde{R}}\partial_{\xi}+\xi+\frac{\pi^2\eta_0^2}{\tilde{R}^2\mathcal{E}}  \end{array}\hspace{-1mm}\right)\left(\hspace{-1mm}\begin{array}{c}
f_{\ell}(\xi) \\
g_{\ell}(\xi)\end{array}\hspace{-1mm}\right)=0.
\end{align} 
Here, the mass term is due to the effective Zeeman field $\frac{\kappa\ell}{\mathcal{E}}$. The first solution of this Dirac equation is:
\begin{align}\label{S13}
f_{\ell}^{(1)}(\xi)&=e^{i\frac{\xi^2\tilde{R}}{4\pi\eta_0}}\hspace{-2mm}\phantom{1}_1F_1\left[i\frac{\kappa^2\ell^2\tilde{R}}{8\pi\mathcal{E}^2\eta_0};\frac{1}{2};-i\frac{\xi^2\tilde{R}}{2\pi\eta_0}\right] ;\\\nonumber
g_{\ell}^{(1)}(\xi)&=\frac{\kappa^2\ell^2\tilde{R}\xi}{2\pi\mathcal{E}^2\eta_0}e^{i\frac{\xi^2\tilde{R}}{4\pi\eta_0}}\hspace{-2mm}\phantom{1}_1F_1\left[1+i\frac{\kappa^2\ell^2\tilde{R}}{8\pi\mathcal{E}^2\eta_0};\frac{3}{2};-i\frac{\xi^2\tilde{R}}{2\pi\eta_0}\right],  
\end{align} 
and the second satisfies $f_{\ell}^{(2)}(\xi)=(g_{\ell}^{(1)}(\xi))^{*}$ and $g_{\ell}^{(2)}(\xi)=(f_{\ell}^{(1)}(\xi))^{*}$. The function $\phantom{1}_1F_1\left[a;b;x\right] $ is the generalized hypergeometric function. 

\begin{figure}[tp]
\begin{flushright}\begin{minipage}{1\textwidth}  \centering
     \includegraphics[width=0.8\textwidth]{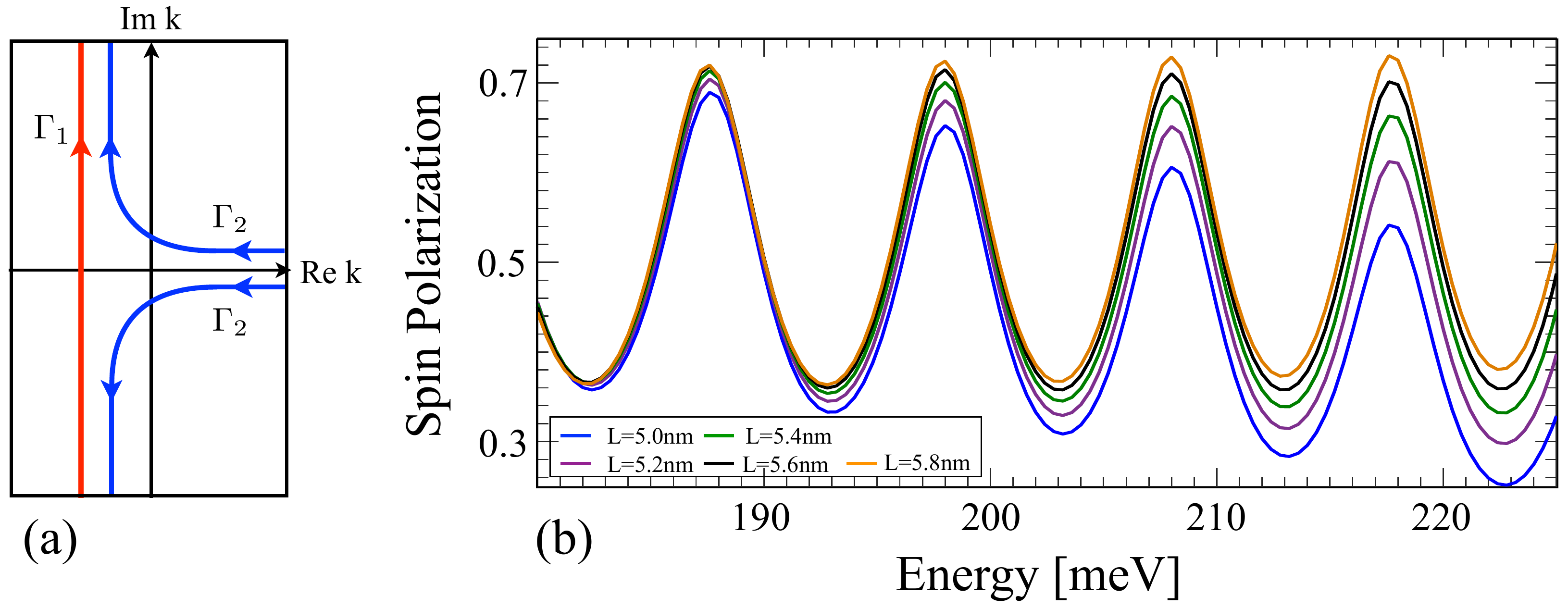} \hspace{0.05in}
              \caption[0.4\textwidth]{\small Spin polarization in the presence of a dipole field. (a) The integration over $k$ in the exact expression for the eigenstates (Eq.~\ref{S14}) is performed along the contour $\Gamma_1$ to obtain the decaying solution, and along $\Gamma_2$ for the diverging wave-function. (b) Changing the molecule length suppresses the transmission, yet increases the spin polarization.}
\end{minipage}\end{flushright}
\end{figure}

Previously we discussed the unique properties of the wave-function in its tail $\xi\gg1$. The asymptotic expansion, however, cannot be easily seen from the expressions for the exact wave-functions given in Eq.~\ref{S13}. To determine the behavior in the tail of the wave-function, it is useful to rewrite $f_{\ell}(\xi)$ and $g_{\ell}(\xi)$ in terms of derivatives. For that purpose, we express the wave-function as an integral over momentum:
 \begin{align}\label{S14}
 \Psi_{\ell}(\xi)=e^{i\gamma\eta_0\ell\xi}\int_{\Gamma}{dk}\left(\begin{array}{c}
f_{\ell}(k) \\
g_{\ell}(k)\end{array}\right)e^{-\frac{k^3}{3}+k\xi}.
\end{align} 
This expression for the wave-function is used to find the decaying (diverging) solution by integrating over the contour $\Gamma_{1}$ ($\Gamma_{2}$) illustrated in Fig.~S2(a). Then, the functions $f_{\ell}(k)$ and $g_{\ell}(k)$ satisfy: 
 \begin{align}\label{S15}
f_{\ell}^{(1)}(k)&=e^{i\frac{\pi\eta_0\xi}{\tilde{R}}}\hspace{-2mm}\phantom{1}_1F_1\left[i\frac{\kappa^2\ell^2\tilde{R}}{8\pi\mathcal{E}^2\eta_0};\frac{1}{2};-i\frac{2\pi\eta_0}{\tilde{R}}\left(k+i\frac{\pi\eta_0}{\tilde{R}}\right)^2\right];\\\nonumber 
 g_{\ell}^{(1)}(k)&=i\frac{\kappa\ell}{\mathcal{E}}\left(k+i\frac{\pi\eta_0}{\tilde{R}}\right)e^{i\frac{\pi\eta_0\xi}{\tilde{R}}}\hspace{-2mm}\phantom{1}_1F_1\left[1+i\frac{\kappa^2\ell^2\tilde{R}}{8\pi\mathcal{E}^2\eta_0};\frac{3}{2};-i\frac{2\pi\eta_0}{\tilde{R}}\left(k+i\frac{\pi\eta_0}{\tilde{R}}\right)^2\right],
\end{align} 
while for the second solution $f_{\ell}^{(2)}(k)=(g_{\ell}^{(1)}(k))^{*}$ and $g_{\ell}^{(2)}(k)=(f_{\ell}^{(1)}(k))^{*}$. Eqs.~\ref{S14} and \ref{S15} allow us to write $\Psi_{\ell}^{(1)}(\xi)$ as:
 \begin{align}\label{S16}
 \Psi_{\ell}^{(1)}(\xi)=e^{i\gamma\eta_0\ell\xi+i\frac{\pi\eta_0\xi}{\tilde{R}}}\left(\begin{array}{c}\hspace{-2mm}\phantom{1}_1F_1\left[i\frac{\kappa^2\ell^2\tilde{R}}{8\pi\mathcal{E}^2\eta_0};\frac{1}{2};-i\frac{2\pi\eta_0}{\tilde{R}}\left(\frac{\partial}{\partial\xi}+i\frac{\pi\eta_0}{\tilde{R}}\right)^2\right] \\
i\frac{\kappa\ell}{\mathcal{E}}\left(\frac{\partial}{\partial\xi}+i\frac{\pi\eta_0}{\tilde{R}}\right)\hspace{-2mm}\phantom{1}_1F_1\left[1+i\frac{\kappa^2\ell^2\tilde{R}}{8\pi\mathcal{E}^2\eta_0};\frac{3}{2};-i\frac{2\pi\eta_0}{\tilde{R}}\left(\frac{\partial}{\partial\xi}+i\frac{\pi\eta_0}{\tilde{R}}\right)^2\right]\end{array}\right)Y(\xi).
\end{align} 

\begin{figure}[]
\begin{flushright}\begin{minipage}{1\textwidth}  \centering
     \includegraphics[width=0.85\textwidth]{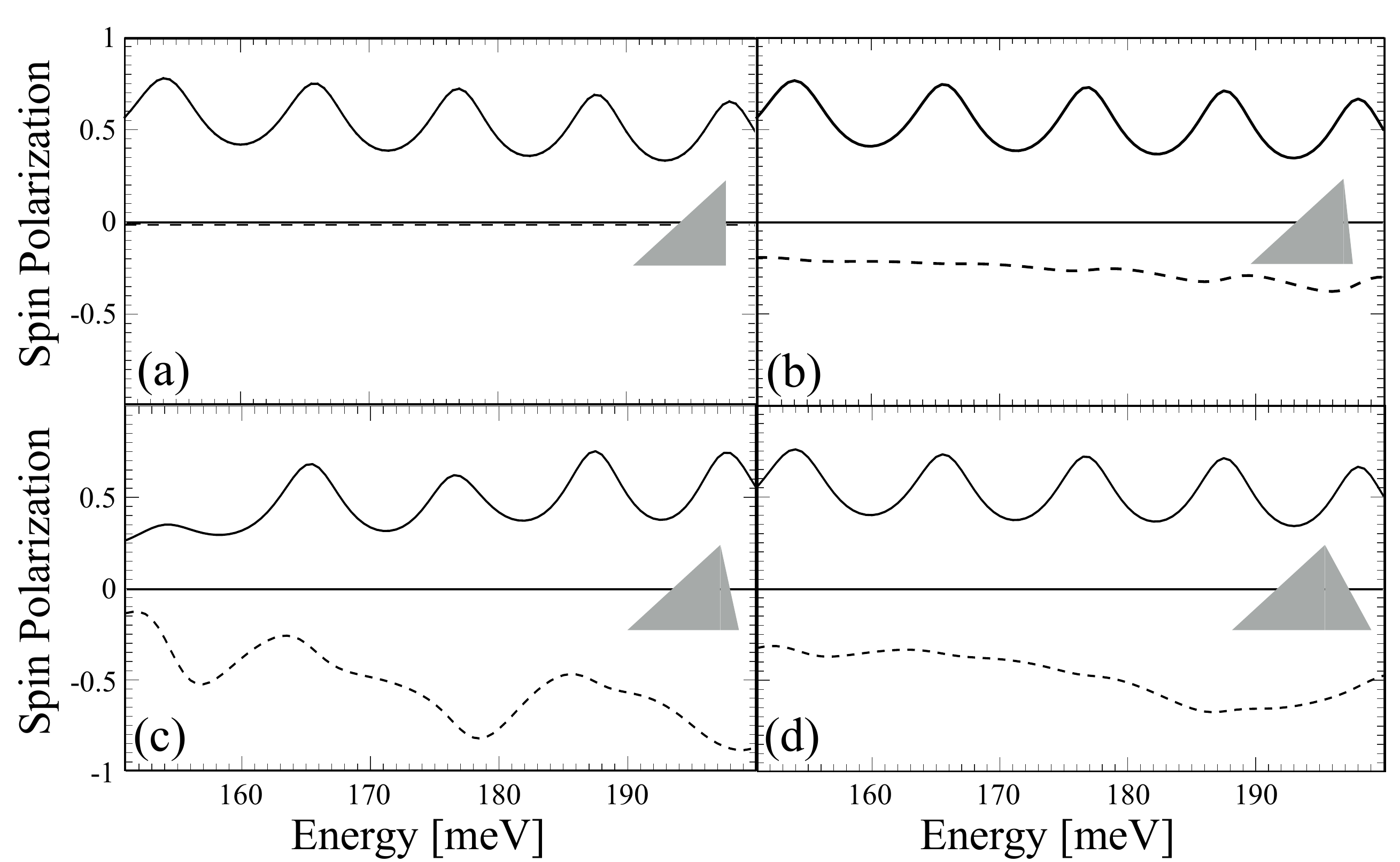} \hspace{0.05in}
              \caption[0.4\textwidth]{\small Spin polarization for current flowing from the left lead to the right (solid), and in the opposite direction (dashed). In all panels the electric potential grows over $5nm$ and drops over a distance of (a) $0nm$ (b) $0.5nm$ (c) $1nm$ and (d) $2.5nm$. The shape of the potential is plotted in gray. }
\end{minipage}\end{flushright}
\end{figure}

Using the relation between $f_{\ell}^{(1)}(\xi)$ and $g_{\ell}^{(1)}(\xi)$ to $f_{\ell}^{(2)}(\xi)$ and $g_{\ell}^{(2)}(\xi)$, we can write a similar expression for $\Psi_{\ell}^{(2)}(\xi)$. To get the asymptotic expansion of the wave-function, we use the fact that for $\xi\gg1$ the derivatives of the Airy functions satisfy $\partial_{\xi}Ai(\xi)\xrightarrow[\xi\gg1]{}-\sqrt{\xi}Ai(\xi)$ and $\partial_{\xi}Bi(\xi)\xrightarrow[\xi\gg1]{}\sqrt{\xi}Bi(\xi)$. Then, from the expansion of the generalized hypergeometric functions for large $\xi$ we get the expression in Eq.~7. The helicity in the tail of the wave-function is found by calculating the momentum per spin:
\begin{align}\label{S17}
\langle\Psi_{\ell}^{(j)}(\xi)|k_{\sigma}|\Psi_{\ell}^{(j)}(\xi)\rangle=\frac{1}{2}\Psi_{\ell,\sigma}^{*}(\xi)\left(-i\partial/\partial{\xi}-\gamma\ell\right)\Psi_{\ell,\sigma}(\xi)-\frac{1}{2}\Psi_{\ell,\sigma}(\xi)\left(i\partial/\partial{\xi}+\gamma\ell\right)\Psi_{\ell,\sigma}^{*}(\xi).
\end{align} 
As we show in .~4(a),  for a right handed helix, the momentum of spin up (down)  is always positive (negative) in the tail for both $\Psi_{\ell}^{(1)}$ and $\Psi_{\ell}^{(2)}$.

The large $\xi$ behavior of the wave-function is written here to demonstrates the unique properties of the electronic states in an helical tube with a large dipole field. To find the spin-dependent scattering matrix through the tube, we have used the exact expression for the eigenstates as given in Eqs.~\ref{S14} and \ref{S15}. Since the calculation demands numerical integration, we were deriving the transmission coefficients only for short molecules of the order of few nanometers. To access a  wider range of lengths, we also performed a tight-binding calculation of the scattering matrix. For this purpose, we consider the discrete version of the Hamiltonian in Eq.~5:
\begin{align}\label{S18}\nonumber
\mathcal{H}&=\sum_{n=1}^{N}\sum_{\sigma=\uparrow,\downarrow}\left[t\left(c_{n+1,\sigma}^{\dag}c_{n,\sigma}+c_{n,\sigma}^{\dag}c_{n+1,\sigma}\right)+E_Dc_{n,\sigma}^{\dag}c_{n,\sigma}\right]\\
&+i\kappa\ell\sum_{n=1}^{N}\left[e^{\frac{2\pi{i}na}{\tilde{R}}}c_{n,\downarrow}^{\dag}c_{n,\uparrow}-e^{-\frac{2\pi{i}na}{\tilde{R}}}c_{n,\uparrow}^{\dag}c_{n,\downarrow}\right].
\end{align} 
Here $c_{n,\sigma}^{\dag}$ ($c_{n,\sigma}$) creates (annihilates) an electron with spin $\sigma$ on site $n$ along the spiral, and the hopping parameter  $t=\frac{\hbar^2}{2ma^2}$ depends on the lattice spacing $a$. Then, we use the method introduced by  Lee and Fisher~\cite{Lee,Giordano}  to obtain the transmission probability. We have shown that the transmission probability as well as the spin polarization calculated in the continuous and discrete models are consistent. Here, we present additional results of the tight binding calculation. First, we present the change in spin polarization as a function of the molecule length. As shown in Fig.~6(b), increasing the length while keeping the electric dipole field constant reduces the transmission, but at the same time enhances the spin polarization. 

Finally, we wish to emphasize that in all the results presented in Fig.~4 we assumed that the dipole field drops sharply to zero at the end of the molecule. This strong asymmetry gives rise to a significant difference in spin-polarization between currents flowing from the left lead to the right and currents going in opposite direction, as shown in Fig.~7(a). Such a sudden drop in the electric field, however, is not physical; a more realistic model should take into account a more gradual decrease of the dipole potential. In Fig.~7 we consider different slopes for the potential drop.  Interestingly, allowing the dipole potential that grows over $5nm$ to decay over $0.5nm$ already increases the spin polarization for electrons transferring from right to left to the same order as the polarization of electrons flowing in the opposite direction.

\end{document}